\documentclass[aps,prd,reprint,superscriptaddress,nofootinbib,floatfix]{revtex4-1}

\usepackage{amsmath}
\usepackage{amssymb}
\usepackage{graphicx}
\usepackage[caption=false]{subfig}
\usepackage{enumitem}
\usepackage{color}
\usepackage{mathtools}
\usepackage[percent]{overpic}
\usepackage{bm}
\usepackage{rotating}
\usepackage{cancel}
\usepackage{hyperref}
\usepackage{wrapfig}
\usepackage{array}
\usepackage[normalem]{ulem}
\usepackage{ifthen} 

\ifthenelse{\isundefined{\showoldabstracts}}{
  \newcommand{\oldabstracts}[1]{}
}{
  \newcommand{\oldabstracts}[1]{#1}
}
\def\shownal{1} 
\newcommand{\nal}[1]{\ifthenelse{\shownal=1}{\textcolor{blue}{[[AL: #1]]}}{}}
\newcommand{\nemma}[1]{\ifthenelse{\shownal=1}{\textcolor{purple}{[[Emma: #1]]}}{}}
\newcommand{\nedu}[1]{\ifthenelse{\shownal=1}{\textcolor{red}{[[Eduardo: #1]]}}{}}

\newcommand{\ket}[1]{{\left| {#1} \right>}}
\newcommand{\bra}[1]{{\left< {#1} \right|}}

\newcommand{\proj}[2]{{\left| {#1} \right>\left< {#2} \right|}}

\newcommand{\ii}{\mathrm{i}}
\newcommand{\ee}{\mathrm{e}}
\newcommand{\tr}{\text{Tr}}
\newcommand{\trpi}{\text{Tr}_{\pi}}

\newcommand{\diff}[1]{\!\!\mathrm{d}{#1}\;}

\begin{document}

\title{Finite sizes and smooth cutoffs in superconducting circuits}

\author{Emma McKay}
\affiliation{Institute for Quantum Computing, University of Waterloo, Waterloo, Ontario, N2L 3G1, Canada}
\affiliation{Department of Applied Mathematics, University of Waterloo, Waterloo, Ontario, N2L 3G1, Canada}

\author{Adrian Lupascu}
\affiliation{Institute for Quantum Computing, University of Waterloo, Waterloo, Ontario, N2L 3G1, Canada}
\affiliation{Department of Physics \& Astronomy, University of Waterloo, Waterloo, Ontario, N2L 3G1, Canada}
\affiliation{Waterloo Institute for Nanotechnology, University of Waterloo, Waterloo, Ontario, N2L 3G1, Canada}

\author{Eduardo Mart\'{i}n-Mart\'{i}nez}
\affiliation{Institute for Quantum Computing, University of Waterloo, Waterloo, Ontario, N2L 3G1, Canada}
\affiliation{Department of Applied Mathematics, University of Waterloo, Waterloo, Ontario, N2L 3G1, Canada}
\affiliation{Perimeter Institute for Theoretical Physics, 31 Caroline St. N., Waterloo, Ontario, N2L 2Y5, Canada}


\begin{abstract}
\oldabstracts{
}
We investigate the validity of two common assumptions in the modelling of  superconducting circuits: first, that the superconducting qubits are  pointlike, and second, that the UV behaviour of the transmission line is not relevant to the qubit dynamics. We show that in the experimentally accessible ultra-strong coupling regime and for short (but attainable) times, the use of an inaccurate cutoff model (such as sharp, or none at all) could introduce very significant inaccuracies in the model's predictions.
\end{abstract}

\maketitle

\section{Introduction}

Implementations of superconducting circuits have made significant advances in recent years, especially those employed in quantum information processing. Among other things, superconducting circuits have provided a platform for the implementation and simulation of most of the features of the light-matter interaction of quantum optics. Remarkably, superconducting circuits can reach coupling regimes far beyond the strength with which atomic systems interact with the electromagnetic field~\cite{Forn-Diaz2010,Yoshihara2016,Niemczyk2010,Fornada}. Furthermore, superconducting qubits can be ultra-strongly coupled to quantum fields in a controlled, time-dependent way~\cite{PeroPadre2010}. In these regimes of fast ultra-strong coupling of light and matter, some of the assumptions usually made in conventional quantum optics may need to be reassessed. 

For example, circuits with (finite-length) resonators have been analyzed using the Jaynes-Cummings model~\cite{Jaynes1963}, which often includes making the rotating wave approximation (RWA) and assuming that the qubit interacts with a single or just a few modes of the electromagnetic field~\cite{Reed2010,Wallraff2004,Schmidt2012,Sabin2010,Lizuan2010,Niemczyk2010}. Ultrastrong coupling research has reached regimes where so-called counter-rotating terms are no longer negligible---thus the RWA is no longer valid~\cite{Fornada,Niemczyk2010,Forn-Diaz2010,Lizuan2010,Sabin2010,Casanova2010,Yoshihara2016,Ashhab2010}. Also, many works in the ultra-strong coupling regime consider only a few modes of the electromagnetic field~\cite{Niemczyk2010,Forn-Diaz2010,Lizuan2010,Sabin2010,Casanova2010,Yoshihara2016}. In contrast, studies of open resonators have delved into analysis of qubit-line coupling over a continuum of modes, and are in the early stages of considering ultrastrong regimes~\cite{Peropadre2013,Fornada,Haeberlein2015}. 

It is therefore relevant to investigate the regime of ultrastrong coupling to a continuum. In particular, this regime opens up analysis of the effect on qubit dynamics of  the mode-dependent interaction strength introduced by the breakdown of Cooper pairs at high frequencies---the natural ultraviolet cutoffs of superconducting materials. Mode-dependence has been studied before in terms of the increase of coupling strength for higher resonant frequencies~\cite{Sundaresan2015}, but it is relevant to also understand  the decay of coupling strength as the frequency of the modes increases.  When frequency cutoffs have been implemented, they are often introduced without a first-principle justification~\cite{Fornada}. In this light, one may wonder whether the specific model of cutoff leads to significant differences in predictions and thus whether it is justified to use an effective cutoff model without deriving it from the microscopic physics of the superconducting setup.

Related to the existence of a UV cutoff is the finite size and nontrivial shape of the qubit. Both the UV cutoff and the physical shape of quantum systems that couple to the  field can have non-trivial effects on the predictions of light-matter interaction models and their causal behavior~\cite{Martin-Martinez2015}. In the particular case of superconducting qubits, the shape of the circuit has not yet been fully considered in the literature as a contributor to observable dynamics. We will also analyze whether the shape of the qubit can contribute to its dynamics in a manner similar to the ultraviolet cutoff. 

As we will see in this paper, short-time dynamics are especially affected by the way in which effective UV cutoffs are implemented. This is particularly interesting since short-time qubit-line interactions are desired for the generation of  so-called single-photon states (i.e. states with low expectation values of the number operator)~\cite{Houck2007,Peng2016}. Additionally, theoretical predictions of states generated through short interaction times have very wide distributions in frequency space, so it is important to take into account how the interaction strength behaves at high frequencies.

In this paper, we will study the system of a qubit and open transmission line in the ultrastrong coupling regime with non-adiabatic switching, taking into account the presence of a natural ultraviolet cutoff in the continuum of modes and the finite size of the qubit. We will use a simple but flexible model to assess the impact of specific forms and scales of qubit shape and ultraviolet cutoff. Specifically, we calculate the probability of vacuum excitation of the qubit and of spontaneous emission into the line, and investigate how different shapes, sizes, cutoff behaviours, and magnitudes of the cutoff scale affect these probabilities. We will show that in the experimentally attainable regime of short interaction, disregarding the behaviour of the cutoff can lead to significant inaccuracies in the model's predictions.

The paper is organized as follows: in section II, we set up our model of a qubit coupled to an infinite 1+1D transmission line and use this to calculate the state of the qubit after an interaction. In section III, we detail the study of how significantly the factors of qubit shape and size and cutoff model and scale impact observable physics of the qubit for different switching times. We show that the cutoff model and scale have a significant impact, which is increased for short and non-adiabatic switchings.

\section{Setup}

\subsection{A simple model for superconducting qubits coupled to a transmission line}

Let us consider a very simple superconducting setup: a superconducting flux qubit coupled to an infinite 1+1D transmission line.

We can identify the following three characteristic parameters of such a setup, summarized in table~\ref{tab:scales}.
The first is the energy gap $\Omega$ between the two energy levels of an idealized superconducting qubit. Typically, superconducting qubits have energy gaps between between 1 and 20 GHz \cite{chiorescu_2004_1,bylander_2011_noisePCQ,yoshihara_2006_1,Stern_2014_3DcavityFluxQubits,orgiazzi_2016_FluxQubitsPlanar}. The second relevant scale is the frequency at which microwave modes no longer propagate through the transmission line. Most common superconductive materials stop behaving as superconductors for electromagnetic frequencies of the order of 100 GHz. More specifically, the propagation of electromagnetic signals in superconducting transmission lines is affected by attenuation, which increases dramatically at frequencies corresponding to the superconducting gap. For aluminum, a material commonly used for superconducting transmission lines, the superconducting gap is 75 GHz~\cite{douglass_1964_Algap}. Experiments in Ref.~\cite{Fornada} showed renormalization of the transition frequency of a qubit coupled to an aluminum transmission line, consistent with a cutoff scale of 50 GHz chosen in this paper, which is close to the value of the superconducting gap for aluminum. In this fashion, there is a natural UV cutoff scale in the problem associated with dissipative effects at higher frequencies. We refer to that UV cutoff scale as $\varepsilon_0$.  The third scale is the physical size of the qubit $\sigma_0$, typically on the order of 10 $\mu$m~\cite{chiorescu_2004_1,bylander_2011_noisePCQ,yoshihara_2006_1,Stern_2014_3DcavityFluxQubits,orgiazzi_2016_FluxQubitsPlanar,Fornada}. In previous literature, this scale is often neglected by considering the qubit to be pointlike in its coupling to the transmission line. However, the finite size of the physical qubit and the long range nature of the interaction of the qubit with the field in the line could, in principle, affect the dynamics of the qubit-line system. Hence, we will not assume a pointlike flux qubit by default in this work.
    
    \begin{table}[ht]        
        \begin{tabular}{l|l|l|l}
        & Physicality               & Approx. value     & Scaled to $\Omega$\\ \hline \hline
    	$\Omega$		& energy gap of qubit		& 10 GHz  	& $\Omega$          \\ \hline
        $\varepsilon_0$	& cutoff scale				& 50 GHz   & $5\,\Omega$        \\ \hline
        $\sigma_0$	    & physical size of qubit	& 10 $\mu$m     & $10^{-4}\,\Omega^{-1}$ 
    	\end{tabular}
    	\caption{Scales of the problem.}
    	\label{tab:scales}
	\end{table}
    
As a theoretical model of the qubit-field interaction, we consider the following interaction Hamiltonian:
    \begin{equation}
        \hat{H}_{\text{int}} = \lambda\; \chi(t)\; \hat{\mu}(t)
        \int\diff{x} F_{\sigma}(x)\;\hat{\pi}(x,t).
        \label{eqn:Hamil}
    \end{equation} 
Here, $\lambda$ is the coupling strength (related to the coupling constant of spin-boson models in appendix~\ref{app:coupling}), $\chi(t)$ is the switching function (analyzed in detail below), \mbox{$\hat{\mu}(t)=\ee^{i\Omega t}\ket{e}\bra{g} + \ee^{-i\Omega t}\ket{g}\bra{e}$} is the qubit's internal degree of freedom, and $F_{\sigma}(x)$ is the spatial distribution of the qubit (which for now we keep arbitrary---it will be made concrete later on in the paper) and the parameter $\sigma$ is the characteristic lengthscale of the qubit. Note that $F_{\sigma}(x)$ describes the `shape' of the qubit and $\sigma$ its `size'---as mentioned above, we can assume these will be related to the physical size of the qubit $\sigma_0$, and could be affected by factors such as the range of the interaction (i.e. the nontrivial decay of the interaction away from the qubit). The observable of the field that the qubit couples to in this simplified model is the conjugate momentum $\hat{\pi}(x,t)$ of a scalar field $\hat\phi$. 
This corresponds to the qubit being coupled to the current in the 1D wire. We can expand $\hat\pi$ in plane waves as
    \begin{equation}
        \hat{\pi}(x,t) = \frac{1}{2\sqrt{\pi}} 
        \int_{-\infty}^{\infty}\!\!\!\!\diff{k} 
        C_{\varepsilon}(k)\sqrt{|k|} 
        \Big(\ii\hat{a}^{\dagger}_k \ee^{\ii(|k|t-kx)} + \text{H.c.}\Big),
    \end{equation}
where $\hat{a}^{\dagger}_k$ and $\hat{a}_k$ are creation and annihilation operators for each mode $k$. This field plays the role of the Ohmic environment in the spin-boson model. See Appendix \ref{app:coupling} for the relationship of this model and the usual spin-boson coupling often used in the literature. Notice that we have included a weight function $C_{\varepsilon}(k)$, which implements a cutoff scale $\varepsilon$. This is done to model the dissipation in higher frequency modes coming from the breakdown of superconductivity and radiative loss. One simple way to visualize the effect of this soft cutoff function is to think that the qubit `sees' the effective coupling strength to different modes monotonically decreasing as the mode frequency increases. Different microscopic models for the dissipation mechanisms would give different shapes and scales for this cutoff function. In a similar manner to $F_{\sigma}(x)$, $C_\varepsilon(k)$ models the `shape' of the UV cutoff model and $\varepsilon$ sets the frequency scale at which the cutoff of higher modes is performed. Note that this cutoff function generalizes the common practice of simply considering a single (or a few) modes in the transmission line, in which case $C_\varepsilon(x)$ would just be a compactly supported function with a sharp decay at $\varepsilon$. 
     
The switching function $\chi(t)$ describes the way we turn the interaction on and off. We will consider a compactly supported switching function (one can think of a switchable coupling for a finite time \cite{PeroPadre2010,Fornada}). We use a smooth switching to avoid extreme non-adiabatic effects \cite{Satz2007,Louko2008}. Additionally, we would like to consider a function that could feasibly be implemented experimentally. To describe such a function, we will employ two time parameters: a ramp-up time $r$---controlling how fast the interaction is switched on and off---and a duration time $T$---controlling how long the interaction strength remains at its maximum value. We assume a switching function of the following form: 
 \begin{equation}
        \chi(t) = 
        \begin{cases}
        \frac12 + \frac12\cos (\pi/r (t + \frac{T}{2}) )
        	& t \in [-\frac{T}{2}-r,-\frac{T}{2})    \\
        1                                                   
        	& t \in [-\frac{T}{2},\frac{T}{2}]      \\
        \frac12 + \frac12\cos (\pi/r (t - \frac{T}{2}) )
        	& t \in (\frac{T}{2},\frac{T}{2}+r]
        \end{cases}.
        \label{eqn:cos-switch}
    \end{equation}
The ramp-up and ramp-down periods are taken to be half periods of a cosine function. From an experimental perspective, any ramping function will have to be constructed digitally through approximately discrete intervals of constant voltage, then transformed into an analog signal through a filter. Thus an experimentally amenable ramping function is one which can be constructed accurately with reasonable long intervals of constant voltage and an achievable analog bandwidth. This is important in view of lower availability of high bandwidth pulse generators and difficulties in preserving pulse shapes due to distortion during transmission from the generator to the switchable coupler between the qubit and transmission line. The cosine ramping fits these requirements. In addition, this pulse shape is more easily amenable to analytic expressions for the transition probabilities than other possible shapes. The particular shape of the switching function is not as important as the presence of two scales, one controlling the adiabaticity of the switching (ramp-up/down time) and another the duration of the interaction. We expect the dependence of the transition probabilities on the ramp-up and ramp-down times to provide insight into experiments with other pulse shapes.     
    \begin{figure}[h!]      
        \centering
        \includegraphics[width=0.35\textwidth]{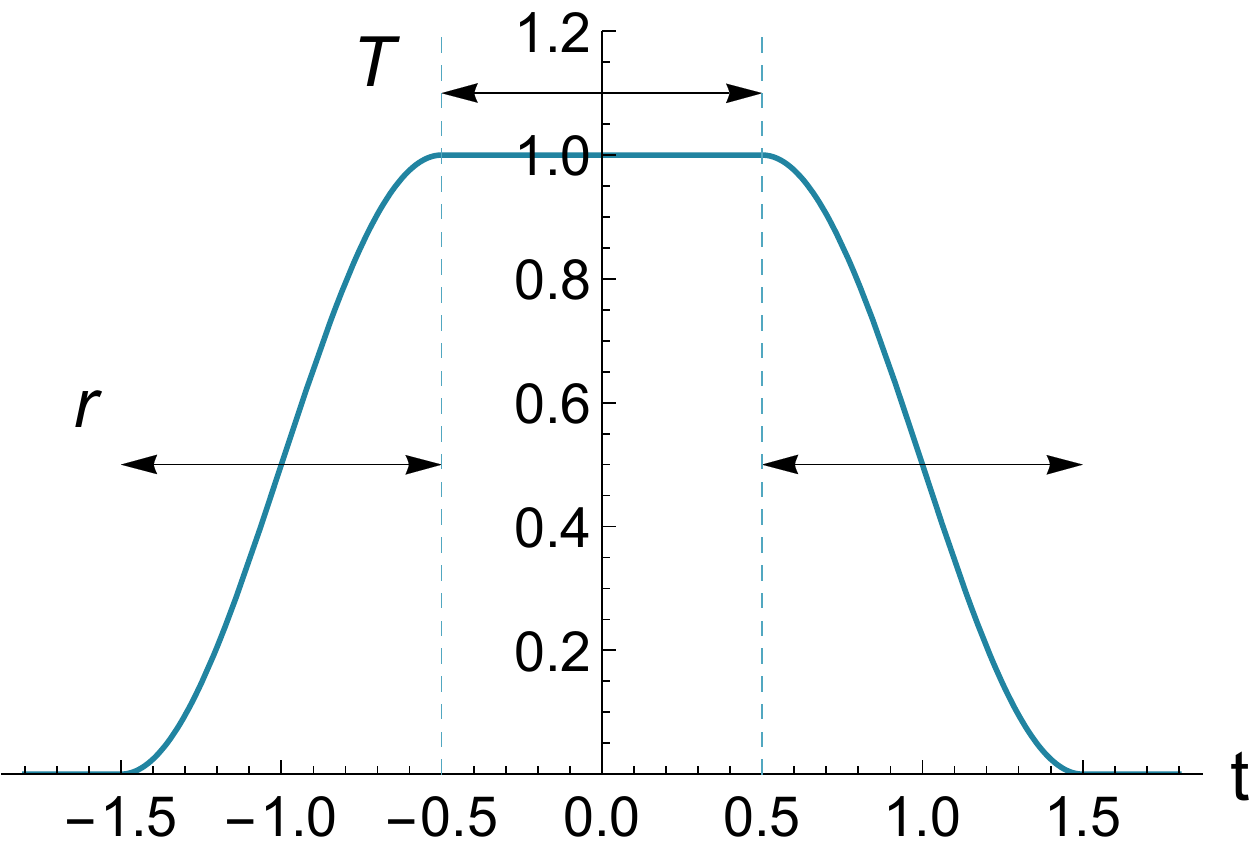}
        \caption{The switching function $\chi(t)$.}
        \label{fig:cos-switch}
    \end{figure}
Both the ramp-up time $r$ and the interaction duration $T$ are tunable and can, in principle, be as short as 0.01 ns in experiment. Recently, a new generation of pulse generators with a time resolution of the order of 10 ps and analog bandwidth of the order of 10 GHz were used for direct digital synthesis of pulses for control of flux~\cite{Deng_2015_Floquet} and transmon qubits~\cite{raftery2017direct}.

In this paper, we will explore a range of both $r$ and $T$ to characterize the behaviour of the model in the short/long and non-adiabatic/adiabatic interaction regimes. Based on previous results on the light matter interaction \cite{Martin-Martinez2015}, we expect the shape of the qubit and the cutoff model to have a more significant effect in the non-adiabatic and short interaction timescales.

\subsection{Time evolution}

We use perturbation theory to calculate the state of the qubit after its interaction with the field as mediated by the Hamiltonian~\eqref{eqn:Hamil} with the cosine trapezoidal switching function~\eqref{eqn:cos-switch}. 

We begin with the field and qubit in their respective free ground states, $\ket{0}$ and $\ket{g}$: 
    \begin{align}
        \hat\rho_\text{in} &= \hat\rho_{\pi,\text{in}} \otimes \hat\rho_{\text{q},\text{in}}= \ket{0}\bra{0} \otimes \ket{g}\bra{g}.
        \label{eqn:initial}
    \end{align}  
Note that we assume that the initial state of the qubit is the ground state for the calculation of vacuum excitation probability below. However, if we let the qubit energy gap $\Omega$ acquire negative values, these computations also give the probability of spontaneous decay to the ground state starting from the excited state for values $\Omega<0$. 

In the interaction picture, the final state after the interaction $\hat\rho_{\text{out}}$ is related to the initial sate $\hat\rho_{\text{in}}$  through the time evolution operator
    \begin{equation}
        \hat U =\mathcal{T}\exp\left(\int_{-\infty}^\infty \diff t \hat H_{I}(t)\right)
        =\mathcal{T}\exp\left(\int_{-T-r}^{T+r} \diff t \hat H_{I}(t)\right),
    \end{equation}
where $\mathcal{T}$ represents the time-ordering operation. Note that $\chi(t)$ has compact support, so the limits of integration can be written equivalently as $[-T-r,T+r]$ or $(-\infty,\infty)$. The Dyson expansion associated with this time evolution operator is given by
    \begin{align} \label{eqn:dyson}
        \hat{U} &= \openone \underbrace{- \,\ii\! \int_{-\infty}^{\infty}\!\!\!\!\diff{t} \hat{H}_I(t)}_{\hat{U}^{(1)}}
        \underbrace{- \!\int_{-\infty}^{\infty}\!\!\!\!\diff{t}\!\! \int_{-\infty}^{t}\!\!\!\!\diff{t'}\! \hat{H}_I(t)\hat{H}_I(t')}_{\hat{U}^{(2)}} +\, \mathcal{O}(\lambda^3).
    \end{align}
Considering this, the qubit-field state after the interaction becomes
    \begin{align}\label{eqn:afterint}
        \hat\rho_{\text{out}} &= \hat{U}\hat\rho_{\text{in}}\hat{U}^{\dagger}= \hat\rho_{\text{in}} + \hat\rho_{\text{out}}^{(1)} + \hat\rho_{\text{out}}^{(2)} + \mathcal{O}(\lambda^3),
    \end{align}
where
    \begin{align}
        \hat\rho_\text{out}^{(1)} &= \hat{U}^{(1)}\hat\rho_{\text{in}} + \hat\rho_{\text{in}}\hat{U}^{(1)\dagger},\\
        \hat\rho_{\text{out}}^{(2)} &= \hat{U}^{(1)}\hat\rho_{\text{in}}\hat{U}^{(1)\dagger} + \hat{U}^{(2)}\hat\rho_{\text{in}} + \hat\rho_{\text{in}}\hat{U}^{(2)\dagger}.
    \end{align}
The state of the qubit after the interaction is obtained by tracing out the field degrees of freedom from $\hat\rho_{\text{out}}$:
    \begin{equation}
        \hat\rho_{\text{q},\text{out}} = \tr_{\pi}[\hat\rho_{\text{in}}] = \hat\rho_{\text{q},\text{in}} + \hat\rho_{\text{q},\text{out}}^{(1)} + \hat\rho_{\text{q},\text{out}}^{(2)}+ \mathcal{O}(\lambda^3).
    \end{equation}
The full detail of these calculations can be found in Appendix~\ref{app:somanycalculations}. For the initial state \eqref{eqn:initial} the first order term vanishes and the second order correction to the initial state yields
    \begin{align}\label{eqn:rhoout}
        \hat\rho_{\text{q},\text{out}}^{(2)}= -P_e \ket{g}\bra{g} + P_e\ket{e}\bra{e}.
    \end{align}
$P_e$ is the leading order contribution to the probability of vacuum excitation of the qubit, and is given by
    \begin{align}\label{eqn:pe}
        \nonumber
        P_e &= \frac{\lambda^2}{4\pi} \int_{-\infty}^\infty \diff{k} \tilde{F}_{\sigma}^2(k) C_{\varepsilon}^2(k) |k|	\\
        & \times\int_{-\infty}^{\infty}\diff{t} \int_{-\infty}^{\infty} \diff{t'}
        \chi(t)\chi(t') \ee^{\ii (|k|+\Omega) (t-t')},
    \end{align}
where 
\begin{equation}\label{eqn:ft}
    \tilde{F}_{\sigma}(k) = \int_{-\infty}^{\infty}\diff{x}F_{\sigma}(x)\ee^{\ii k x}
\end{equation} is the Fourier transform of the qubit's spatial smearing.

Although the spatial smearing $\tilde F_{\sigma}(k)$ (associated with the physical shape, size, and interaction range of the qubit) and the cutoff function $C_\varepsilon(k)$ (associated with the effective cutoff imposed by the dissipation for high frequencies) are very different in origin, they affect the excitation probability on equal footing. In view of this, we can regard the product   $\tilde{F}^2_{\sigma}(k) C^2_{\varepsilon}(k)$ as an \textit{effective form factor} of the qubit.

The $t$ and $t'$ integrals can be expressed in closed form. Substituting in the cosine trapezoidal switching function \eqref{eqn:cos-switch}, we get the following expression for $P_e$:
    \begin{align}\label{eqn:vac-excitation}
        P_e &= \frac{\lambda^2}{4\pi} \int_{-\infty}^{\infty} \diff{k} \tilde{F}_{\sigma}^2(k)\: C_{\varepsilon}^2(k)\, |k| \\ \nonumber
        \times & -\frac{\ii\pi^4}{2} \frac{\ee^{-\frac{1}{2} \ii (k+\Omega ) (2 r+T)}}{(k + \Omega)^{2} (\pi - r(k + \Omega))^{2} (\pi + r(k + \Omega)^{2}} \\ \nonumber
        \times & \left(1+\ee^{\ii r (k+\Omega )}\right)  \left(-1+\ee^{\ii (k+\Omega ) (r+T)}\right) \\ \nonumber
        \times & \left[\sin \left(\frac{1}{2} (k+\Omega ) (2 r+T)\right)+\sin \left(\frac{1}{2} T (k+\Omega )\right)\right].
    \end{align}
Notice that the probability of de-excitation of a qubit initialized in the excited state (i.e. spontaneous emission) is given by the same expression \eqref{eqn:vac-excitation} once we perform the substitution $\Omega\to-\Omega$.

\section{Results and Discussion}

We analyze the effects on the qubit dynamics of variations of shape and size of the qubit and of different cutoff models and scales. In particular, we will characterize the impact of the effective form factor of the qubit on the probability of vacuum excitation and spontaneous emission \eqref{eqn:vac-excitation} in terms of the interaction ramp-up time $r$ and the interaction duration $T$.

For the shape of the qubit we consider four different models---Gaussian, Lorentzian, quartic decay, and sharp decay---and a range of sizes for each. For the specific form of these shape functions, see table~\ref{tab:spatial}. As for the cutoff model, we also consider four different models---Gaussian, Lorentzian, exponential, and sharp---and a range of cutoff scales, including asymptotic behaviour. Notice that we assume that there is no cutoff weight for frequencies below the qubit gap. For the specific form of these UV-cutoff functions, see table~\ref{tab:cutoffs}. 
    \begin{table}[ht]        
        \centering
        \begin{tabular}{m{0.12\textwidth}|m{0.07\textwidth}|m{0.2\textwidth}}
        Smearing & Notation & $\hspace{1.2cm}F_\sigma(x)$\\
        \hline  \hline
            Gaussian & $\:F_{\textsc{g},\sigma}(x)$ &
            \vspace{-0.25cm}
            \begin{equation*}
                \frac{1}{\sigma\sqrt{\pi}} \ee^{-x^2/\sigma^2}
            \end{equation*} 
            \vspace{-0.25cm} \\ \hline
            Lorentzian & $\:F_{\textsc{l},\sigma}(x)$ &
            \vspace{-0.2cm}
            \begin{equation*}
                \frac{\sigma}{2\pi} \frac{1}{x^2+(\sigma/2)^2}
            \end{equation*}
            \vspace{-0.25cm}\\ \hline
            Quartic decay & $\:F_{\textsc{q},\sigma}(x)$ &
            \vspace{-0.4cm}
            \begin{equation*}
                \frac{\sqrt{2}(\sigma/2)^3}{\pi} \frac{1}{x^4+(\sigma/2)^4}
            \end{equation*}
            \vspace{-0.4cm}\\ \hline
            Sharp decay & $\:F_{\textsc{s},\sigma}(x)$ & 
            \vspace{-0.4cm}
            \begin{equation*}
                \frac{1}{\sigma} \Theta(x+\sigma/2,-x+\sigma/2)
            \end{equation*}
            \vspace{-0.4cm}
        \end{tabular}
        \caption{Spatial distributions.}
        \label{tab:spatial}
    \end{table}

    \begin{table}[ht]        
    \centering
    \begin{tabular}{m{0.1\textwidth}|m{0.07\textwidth}|m{0.22\textwidth}}
    Cutoff & Notation & $\hspace{1.2cm}C_\varepsilon(x)$\\
        \hline  \hline
        Gaussian & $\:C_{\textsc{g},\varepsilon}(k)$ & \vspace{-0.35cm}
            \begin{equation*}
            \begin{cases}
                \ee^{-(|k|-\Omega)^2/2\varepsilon^2}& |k| > \Omega \\
                1   & |k| < \Omega
            \end{cases}
            \end{equation*} 
            \vspace{-0.25cm} \\ \hline
        Lorentzian & $\:C_{\textsc{l},\varepsilon}(k)$ & \vspace{-0.3cm}
            \begin{equation*}
            \begin{cases}
                \frac{\varepsilon^2}{(|k|-\Omega)^2+\varepsilon^2}& |k| > \Omega \\
                1   & |k| < \Omega
            \end{cases}
            \end{equation*}  
            \vspace{-0.15cm}\\\hline
        Exponential & $\:C_{\textsc{e},\varepsilon}(k)$ & \vspace{-0.35cm}
            \begin{equation*}
            \begin{cases}
                \ee^{(-|k|+\Omega)/\sqrt{2}\varepsilon} & |k| > \Omega \\
                1   & |k| < \Omega
            \end{cases}
            \end{equation*}
            \vspace{-0.25cm}\\ \hline
        Sharp & $\:C_{\textsc{s},\varepsilon}(k)$ & \vspace{-0.5cm} 
            \begin{equation*}
                \Theta(k+\Omega+\varepsilon,-k+\Omega+\varepsilon)
            \end{equation*}
            \vspace{-0.6cm}
    \end{tabular}
    \caption{Cutoff models.}
    \label{tab:cutoffs}
    \end{table}

The analysis is conducted by calculating the relative difference, referred to as $\Delta_{\textsc{a}\textsc{b}}$, between shape or cutoff models A and B over a range of sizes, cutoff scales, ramp-up times, and interaction times:
    \begin{equation}
        \Delta_{\textsc{a}\textsc{b}} = \frac{|P_{e,\textsc{a}} - P_{e,\textsc{b}}|}{\max \left( P_{e,\textsc{a}}, P_{e,\textsc{b}}\right)}.
    \end{equation}
where A and B denote which models are being compared, e.g. the relative difference between the exponential and sharp cutoffs is denoted as $\Delta_{\textsc{e}\textsc{s}}$. 

The relative differences $\Delta_{\textsc{a}\textsc{b}}$ quantify just how much each model affects the probability of vacuum excitation or spontaneous emission, and in particular estimates the error generated by analyzing data using an incorrect model. Notice that, since $\Delta_{\textsc{ab}}$ is a relative difference, it is independent of the magnitude of the coupling strength $\lambda$.

In the following subsections, we assess the effect on vacuum excitation and spontaneous emission of the components of the effective form factor (shape and cutoff model) and of the two time scales of the interaction (ramp-up time and duration). In section~\ref{sec:models}, we detail how the cutoff model dominates over the shape model in the effective form factor. In section~\ref{sec:scales}, we show that the effect of the size is negligible, while the effect of the cutoff scale is significant. In section~\ref{sec:times}, we detail the dependence of the cutoff model on the length of the interaction time and the ramp-up time of the switching, and how this dependence differs between excitation and emission.

\subsection{Sensitivity to the shape model and cutoff model}\label{sec:models}
    
    \begin{figure*}     
        \centering
        \includegraphics[width=\textwidth]{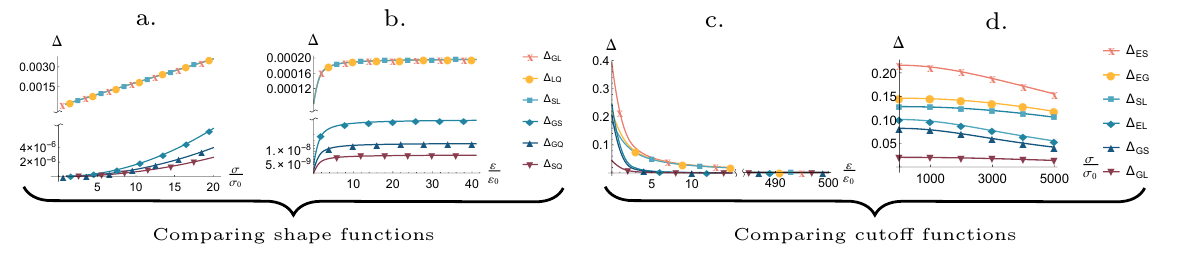}
        \caption{(Colour online). Relative difference of spontaneous emission probabilities when comparing 
        \textbf{a)} shape functions over a range of sizes $\sigma$, 
        \textbf{b)} shape function over a range of cutoff scales $\varepsilon$, 
        \textbf{c)} cutoff functions over a range of cutoff scales $\varepsilon$, and
        \textbf{d)} cutoff functions over a range of sizes $\sigma$.
        Markers are included to distinguish lines. Markers indicate which pair of shape functions or cutoff functions are being compared, and are listed in the legends to the right of figures \textbf{b)} and \textbf{d)}.
        The interaction times are fixed at $r=r_0$ and $T=T_0$. Shapes are compared using exponential cutoff and cutoffs using Gaussian shape. Sizes are compared while the cutoff is kept constant at $\varepsilon_0$ and cutoff scales while the size is kept constant at $\sigma_0$. The behaviour of the vacuum excitation probabilities is similar.}
        \label{fig:emit-sig-eps}
    \end{figure*}

In this section we compare the influence of the cutoff model and the qubit's shape model on the probability of vacuum excitation and spontaneous emission over a range of sizes and cutoff scales, for fixed interaction time scales $r=r_0$ and $T=T_0$. We refer to results displayed in Fig. \ref{fig:emit-sig-eps} for probability of spontaneous emission; the results for vacuum excitation are similar.

Remarkably, the difference effected by the choice of cutoff model is at least two orders of magnitude larger than that effected by the choice of shape for cutoff scales up to at least $20\varepsilon_0$, as seen by comparing panels b and c of Fig.~\ref{fig:emit-sig-eps}. Thus, for all cutoff scales within at least an order of magnitude of $\varepsilon_0$, the choice of cutoff model has a much larger impact on the qubit dynamics than the choice of shape. The cutoff model dominating over the shape model persists for a range of sizes $\sigma$ within several orders of magnitude of the qubit's physical size $\sigma_0$, shown by comparison of panels a and d of Fig.~\ref{fig:emit-sig-eps}.

The dominance of the cutoff model is expected from Eq.~\eqref{eqn:vac-excitation} for emission/excitation probability. The Fourier transform of the shape $\tilde{F}_{\sigma}(k)$, as in Eq.~\eqref{eqn:ft} and the cutoff model $C_{\varepsilon}(k)$, as noted earlier, contribute equally to the $k$ integrand. The scale of the cutoff is, however, about 4 orders of magnitude larger than that of the shape. We thus anticipate that changes to the cutoff model will have a much larger impact on qubit dynamics than changes to the shape model.

Thus the effect of the shape function on  the qubit dynamics is negligible as compared with the effect of the cutoff model. Given this irrelevance of the qubit shape for the regime studied, for the rest of this investigation, we choose the shape to be Gaussian for purely aesthetic reasons.  

\subsection{Sensitivity to size and cutoff scale}\label{sec:scales}

In this subsection we focus on assessing the impact on the qubit dynamics of the physical size of the qubit $\sigma$ and the cutoff scale $\varepsilon$. We again analyze the processes of spontaneous emission and vacuum excitation, when the finite time scales of the interaction are fixed. 

It can be seen in Fig.~\ref{fig:emit-sig-eps}d that the relative difference between different cutoff models is effectively insensitive to the size of the qubit $\sigma$---in fact, a relative difference of just $10^{-6}$ is seen between sizes of $0.1\sigma_0$ and $10\sigma_0$. Thus, as long as the size of the qubit as seen by the transmission line $\sigma$ is within a couple of orders of magnitude of the physical size of the qubit $\sigma_0$, the specific value will have a negligible effect on the dynamics. In view of the fact that the effective form factor of the qubit is largely independent of $\sigma$ for a wide range of values, we will set $\sigma=\sigma_0$ from now on.

In contrast, Fig.~\ref{fig:emit-sig-eps}c shows high sensitivity to the cutoff scale---a relative difference of $0.7$ is seen between cutoff scales of $0.1\varepsilon_0$ and $10\varepsilon_0$.

Note that the relative difference between pairs of cutoff models $\Delta_{\textsc{AB}}$ decreases as $\varepsilon$ increases. This is of course as expected, i.e. as the cutoff scale goes to infinity, the probabilities calculated with different cutoff models converge to the cutoff-free value. Observe here that $P_e$ with an exponential cutoff is slower to converge than the other models considered. Unlike the Gaussian or the Lorentzian models, the exponential cutoff function suppresses the lower frequencies at a higher rate than the Gaussian and Lorentzian cutoffs (which present inflection points at $\varepsilon+\Omega$) or the sharp cutoff. 

From this set of observations we can conclude that for the two scales of the effective form factor, the specific value of the size is irrelevant to qubit dynamics in comparison to the cutoff scale.
    
We can thus say that, for the parameter regimes in table \ref{tab:scales}, the effective form factor of the qubit is dominated by the cutoff model and cutoff scale, and that the effects of the shape model and size are negligible. Intuition for this can be drawn from the difference of several orders of magnitude of the cutoff scale $\varepsilon_0\sim 5\Omega$ and the physical size of the qubit $\sigma_0\sim10^{-4}\Omega^{-1}$. This corroborates the current practice of treating the shape of the qubit as negligible.

\subsection{Sensitivity to cutoff models as a function of the switching time scales}\label{sec:times}    
    \begin{figure}          
        \centering
        \includegraphics[width=0.45\textwidth]{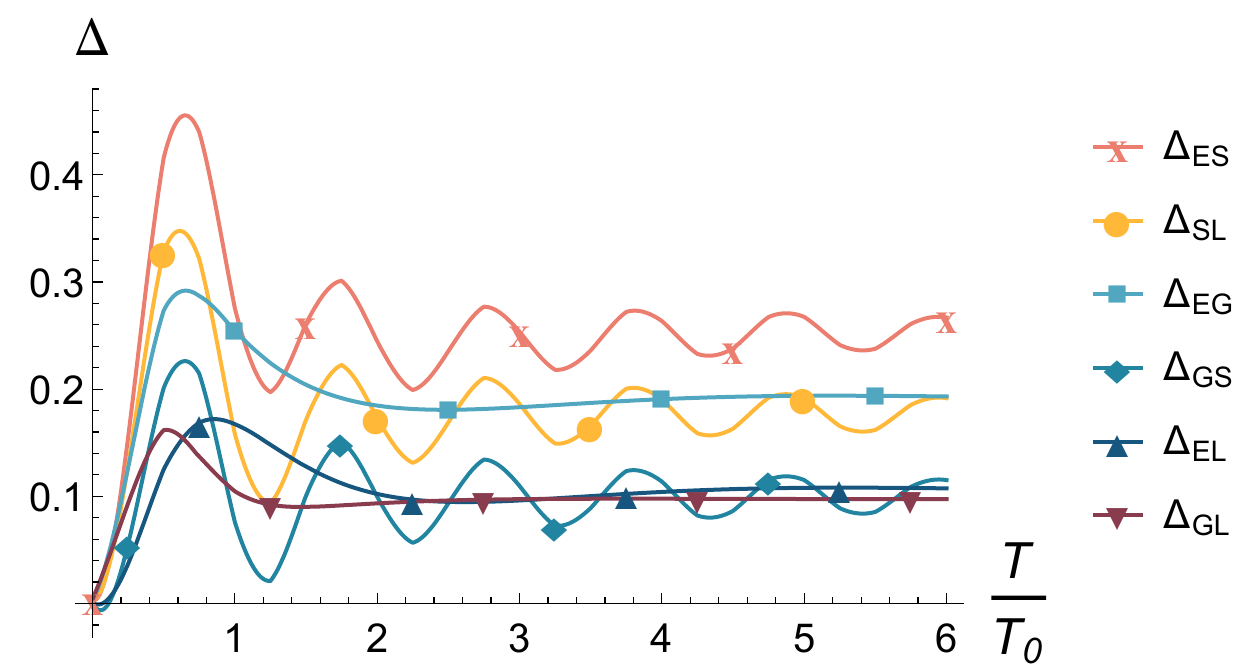}
        \caption{(Colour online.) Relative difference of vacuum excitation probability between pairs of cutoff functions across a range of interaction time scales $T$, for fixed size $\sigma_0$, shape (Gaussian), cutoff scale $\varepsilon_0$ and ramp-up time $r=0.1r_0$. Markers are included to distinguish lines.}
        \label{fig:largeT}
    \end{figure}

We have now established that for typical superconducting qubit setups in the USC regime (see table \ref{tab:scales}), the cutoff model and cutoff scale dominate the effective form factor of a qubit interacting with a superconducting transmission line. As discussed above, superconducting qubits and transmission lines cannot support arbitrarily high frequency modes. This can be traced back to the break down of superconductivity (effecting dissipation in the transmission line) at frequencies above the superconducting gap. The cutoff scale can thus be established, but the exact functional form of the UV cutoff function in a realistic scenario is complex to obtain from first principles, as it involves the complicated interplay of electrodynamics and quasi-particle physics. 

The question then arises when we make predictions as to how much we should care about the particular way in which the transmission line loses its ability to trap higher frequency modes. In other words, will an experiment be sensitive to the particular shape of the cutoff function or just its scale? How much do the microscopics of the superconductor impact the outcome of experiments? Does it really matter if the effective coupling strength decays exponentially with the mode frequency or with any other shape?

    \begin{figure*}     
        \centering
        \includegraphics[width=0.9\textwidth]{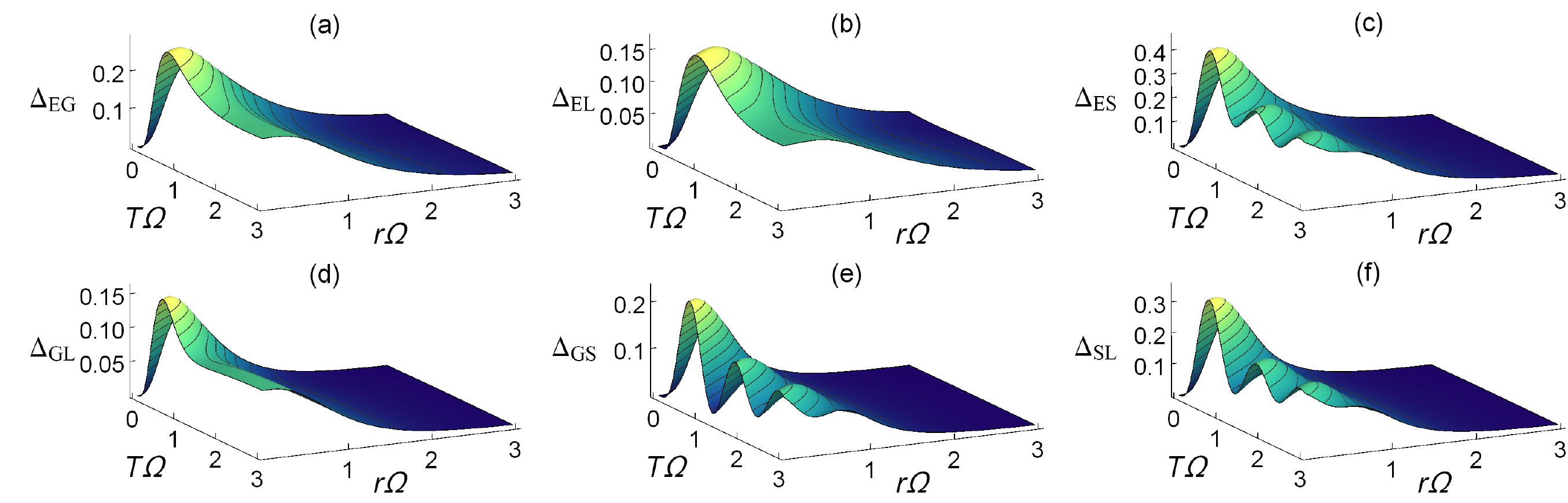}
        \caption{(Colour online.) Relative difference of vacuum excitation probability comparing a) exponential and Gaussian, b) exponential and Lorentzian, c) exponential and sharp, d) Gaussian and Lorentzian, e) Gaussian and sharp, and f) sharp and Lorentzian cutoff functions across a range of ramp up and constant interaction times $r$ and $T$.}
        \label{fig:excite-rT}
    \end{figure*}
    
    \begin{figure*}     
        \centering
        \includegraphics[width=0.9\textwidth]{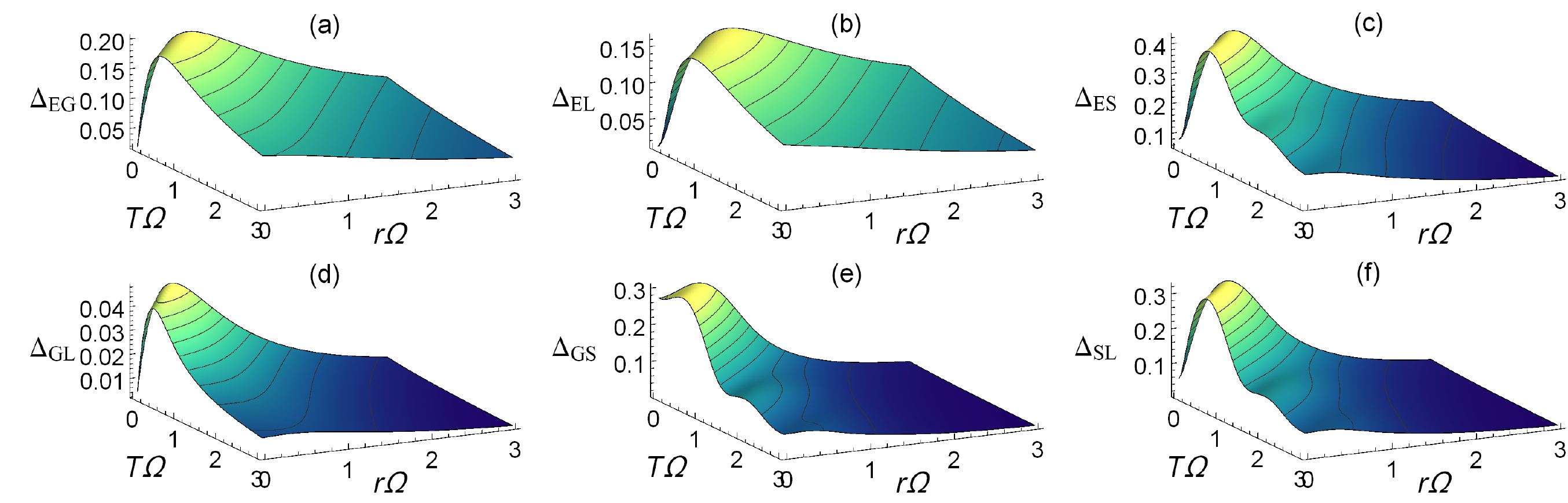}
        \caption{(Colour online.) Relative difference of probability of spontaneous emission comparing a) exponential and Gaussian, b) exponential and Lorentzian, c) exponential and sharp, d) Gaussian and Lorentzian, e) Gaussian and sharp, and f) sharp and Lorentzian cutoff functions across a range of ramp up and constant interaction times $r$ and $T$.}
        \label{fig:emit-rT}
    \end{figure*}

To answer these questions, we are going to assess the impact of the shape of the cutoff function on superconducting experiments with fast finite-time switchable coupling. For this, we will take $\varepsilon=\varepsilon_0$ as a given quantity and explore only the effect of cutoff model. More concretely, we will analyze how the cutoff function influences the probability of vacuum excitation and spontaneous emission for different switching times. Recall that the switching is described by~\eqref{eqn:cos-switch}, with two very different time parameters: the ramp-up time $r$, describing the adiabaticity of the switching process, and the interaction time $T$, describing the length of constant interaction time between switching on and switching off.

The relative difference in the transition probability (both for spontaneous emission and vacuum excitation) between pairs of cutoff models is plotted as a function of $r$ and $T$ in Fig.~\ref{fig:excite-rT} and~\ref{fig:emit-rT}. Here we note a difference in behaviour between spontaneous emission and vacuum excitation processes: in emission, the ramp-up and constant interaction time have similar impacts, while in excitation, the ramp-up time is much more important than the constant interaction time. In Fig.~\ref{fig:largeT}, we see that in vacuum excitation, if the switching is sudden ($r\lesssim \Omega^{-1}$), having the interaction switched on for a long time $T\gg\Omega^{-1}$ does not remove the model dependence of the prediction. In other words, the impact of the cutoff model on excitation is dictated by the switching speed: whether the prediction is sensitive to the particular cutoff model is not dependent on how long the interaction is on, but rather on how suddenly the interaction was switched. In particular, for values of $r\approx \Omega^{-1}$, the relative difference becomes very significant even for large values of the interaction duration $T$ (of the order of $15\%-40\%$ for the parameter values in table~\ref{tab:scales}). Indeed we see that in the adiabatic limit $r\gg\Omega^{-1}$, the prediction is insensitive to the particular cutoff model. If we were trying to anticipate the behaviour of a superconducting qubit coupled to a transmission line in the adiabatic regime we could use any model that facilitates calculations without having to worry about the microscopic mechanism of the dissipation process. 

In comparison, in either the adiabatic or long interaction regimes, the dependence of predictions of spontaneous emission on the cutoff model is negligible. It is only for adiabatic and short interactions that the microscopic mechanisms of dissipation of high frequency modes makes a difference in the predictions.

A few experimental aspects have to be considered for the implementation of fast switching times in superconducting circuits. The fastest arbitrary waveform generators can output switching waveforms with ramp-up and ramp-down times of the order of 20 ps. Other types of pulse generators have even shorter rise/fall times, but lack control of the pulse shape. Propagation of the pulses from room temperature equipment to a tunable coupler between the qubit and the transmission line is affected by distortion, due to dispersion and reflection of pulses at interconnects. Experiments in Ref.~\cite{Deng_2015_Floquet} indicated that pulses could be transmitted between a fast arbitrary generator and a flux qubit with less than 100 ps of edge distortion. Reduction of pulse distortion effects significantly below 100 ps may be possible with significant investments in customized setups. Finally, one needs to consider that coupling elements are usually assumed to work in the adiabatic regime. These coupling elements usually have transition frequencies of the order of tens of GHz and therefore their design has to be optimized to reduce non-adiabatic transitions.

We would like to highlight that the values of $r$ and $T$ for the observation of relevant non-adiabatic dependence of the qubit response on the effective shape of the qubit (and therefore the specific form of the cutoff function) are well within reach of current superconducting qubit technology. A poor choice of cutoff model in these regimes can introduce errors in the theoretical analysis of the setup that are comparable to or larger than other sources of errors that are not neglected in previous analysis.

\section{Conclusion}

In this work, we have introduced a simple model for an ideal superconducting circuit consisting of a flux qubit coupled to an infinite 1+1D transmission line. We specifically address the physics of ultrastrong coupling of a qubit to a continuum of modes by accounting for the mode dependence of superconducting behaviour, the finite size of the qubit, and the long-range interaction between the qubit and transmission line.

We have investigated the effect of the shape model, size, frequency dependence of the coupling strength (cutoff model), and cutoff scale on the probability of the qubit undergoing spontaneous emission and vacuum excitation. We have determined that the cutoff scale and model are the dominant factors in determining the effective form factor of the qubit, while the physical size and shape of the qubit do not make a noticeable difference in comparison. This is consistent with the common practice of not considering the shape of qubits as an important factor in superconducting circuit models.

Concretely, we have analyzed the effect of the coupling strength decaying with frequency (i.e. a cutoff model) in experiments with finite-time switching of the interaction. We have found that, if the switching process is fast (short-lived) and non-adiabatic (rapidly-switched), the cutoff model has a very significant effect on the probability of both vacuum excitation and spontaneous emission. This effect can be comparable to or larger than other sources of error in the experiment. The relative difference in observable quantities between several cutoff models can even be of the order of $10\%$ in experimentally attainable regimes. For long interaction times, the probability of spontaneous emission becomes insensitive to the cutoff model. However, for vacuum excitation, the large differences can remain for arbitrarily long times if the switching ramp-up is non-adiabatic (i.e., the maximum interaction strength is reached in times comparable to the inverse of the qubit gap, which again is experimentally feasible~\cite{Deng_2015_Floquet}). We conclude therefore that assuming a particular cutoff model (e.g. ignoring all frequencies of the field above a specified cutoff) without being careful about the specific way in which this UV cutoff is implemented may lead to inaccurate predictions with these kinds of models with fast switchings and, depending of the particular process studied, for long evolution times.

To end with a bit of a philosophical note, it is perhaps interesting to think of our result in terms of relational ontology: in this work we have arrived to an effective form factor in the qubit-line interaction which constitutes a shape that emerges from the particularities of the interaction. In other words, the shape of the qubit cannot be determined just with an individual description of the qubit itself. Rather, this shape belongs neither to the qubit nor to the line but to the both of them in interaction with each other, constituting a property that becomes evident and relevant in and through interactions between the relevant quantum systems. For further discussion on how properties of quantum systems are difficult to individualize, see, for instance,~\cite{Barad2007}.

\appendix 
\section{Relation of coupling constant to the spin-boson model}\label{app:coupling}

In this appendix we relate the coupling constant $\lambda$ of our model~\eqref{eqn:Hamil} to that of the spin-boson model $\alpha_{\text{SB}}$, as used in Ref.~\cite{Fornada}. 

We start by considering the interaction Hamiltonian $\hat H_{\text{int}}$ in the Schr\"odinger picture given in the Supplementary Information of Ref.~\cite{Fornada}:
    \begin{equation}\label{eqn:spinboson}
    \hat H_{\text{int}} = \hat\sigma_x \sum_k g_k (\hat b_k + \hat b^{\dagger}_k),
    \end{equation}
where $\hat\sigma_x$ is the qubit Pauli operator expressed in the energy eigenbasis and we sum over all bosonic modes. $\hat b_k$ and $g_k$ are respectively the annihilation operator and the coupling constant for mode $k$. In the language of the spin-boson model, the field, which is  acting as the environment to the qubit, is characterized by the spectral density function
    \begin{equation}\label{eqn:spectraldensity}
        J(\omega)= \frac{2\pi}{\hbar^2} \sum_k g^2_k \: \delta(\omega-\omega_k),
    \end{equation}
as in Eq. (35) of the SI of~\cite{Fornada}, where $\omega_k = c\,k$ is the frequency of mode $k$, with $c$ the speed of light. In 1+1D, the spectral density of the electromagnetic field is Ohmic, that is, the power is proportional to the frequency:
    \begin{equation}\label{eqn:ohmicdensity}
        J(\omega)=\pi\omega\alpha_{\text{SB}}
    \end{equation}
Equating spectral densities~\eqref{eqn:spectraldensity} and~\eqref{eqn:ohmicdensity} gives
    \begin{equation}
        g_k = \frac{\hbar}{\sqrt{2}} \sqrt{\alpha_{\text{SB}}\, \omega_k \, \Delta \omega_k}
    \end{equation}
where $\Delta \omega_k$ is the frequency difference between neighbouring modes. Writing this expression for $g_k$ into Eq.~\eqref{eqn:spinboson} gives
    \begin{equation}
        \hat H_{\text{int}} = \hat\sigma_x \sum_k \Delta\omega_k \frac{\hbar}{\sqrt{2}} \sqrt{\alpha_{\text{SB}} \omega_k} \left(\frac{\hat b_k}{\Delta \omega_k} + \frac{\hat b^{\dagger}_k}{\Delta \omega_k} \right).
    \end{equation}
Taking the continuum limit of $\Delta \omega_k\rightarrow\, \diff{\omega_k}$ allows us to write
    \begin{equation}\label{eqn:spinboson2}
        \hat H_{\text{int}} = \hat\sigma_x \frac{\hbar}{\sqrt{2}} \int_{-\infty}^{\infty} \diff{\omega_k}  \sqrt{\alpha_{\text{SB}} \omega_k} \, (\hat a_k + \hat a^{\dagger}_k).
    \end{equation}
where the $\hat a_k$ are new annihilation operators, now defined over a continuum of modes $k$. Now, we can directly compare this with our Hamiltonian~\eqref{eqn:Hamil}, recalling a few facts: our Hamiltonian is written in the interaction picture; our qubit's monopole moment $\hat\mu(t)$ is given by $\hat\sigma_x$; $\omega_k = c\,k$; and in the body of this paper, we have taken $\hbar=c=1$. We can thus write the coupling constant $\lambda$ of Eq.~\eqref{eqn:Hamil} in terms of quantities of the spin-boson model~\eqref{eqn:spinboson2} as
    \begin{equation}
        \lambda = \sqrt{2\pi\,\alpha_{\text{SB}}}.
    \end{equation}
We end this appendix by noting that in Ref.~\cite{Fornada}, the spin-boson coupling parameter $\alpha_{\text{SB}}$ reached values of order 1, indicating that the coupling constant $\lambda$ can also reach values of order 1.

\section{Calculation of the state}\label{app:somanycalculations}

In this appendix we detail the calculation of the state of the qubit as in equation~\eqref{eqn:rhoout} after interaction with the field as described by the Hamiltonian~\eqref{eqn:Hamil}. 

We treat the interaction Hamiltonian as a perturbation to the qubit free dynamics, using the Dyson expansion~\eqref{eqn:dyson} to calculate the first and second order evolution operators $\hat U^{(1)}$ and $\hat U^{(2)}$. The state of the qubit after the interaction is given by tracing out the field from the state $\hat\rho_{\text{out}}$ as in equation~\eqref{eqn:afterint}:
    \begin{equation}
        \hat\rho_{\text{q},\text{out}} = \tr_{\pi}[\hat\rho_{\text{in}}] = \hat\rho_{\text{q},\text{in}} + \hat\rho_{\text{q},\text{out}}^{(1)} + \hat\rho_{\text{q},\text{out}}^{(2)}+ \mathcal{O}(\lambda^3).
    \end{equation}
If the field's initial state is the vacuum, the one-point function  $\tr_{\pi} ( [\hat\rho_{\pi,0},\hat{\pi}(x,t)])=0$. This means that the leading order contribution to the qubit dynamics will be given by terms of order $\mathcal{O}(\lambda^2)$. 

We can thus begin by writing out the second order term in full:
    \begin{align} \nonumber
        \hat\rho_{\text{q},\text{out}}^{(2)} &= - \lambda^2 \trpi \left(\int_{-\infty}^{\infty}\diff{t} \int_{-\infty}^{\infty}\diff{t'} \chi(t)\chi(t') \hat{\mu}(t)\hat\rho_{\text{q},0}\hat{\mu}(t')\right.\\ \nonumber &\int\diff{x}\int\diff{x'}F_{\sigma}(x)F_{\sigma}(x') \hat{\pi}(x,t)\proj{0}{0}\hat{\pi}(x',t') \\
        + &\int_{-\infty}^{\infty}\diff{t}\int_{-\infty}^t\diff{t'}
        \chi(t)\chi(t')\hat{\mu}(t)\hat{\mu}(t')\hat\rho_{\text{q},0}\\ \nonumber
        &\left.\int\diff{x}\!\!\int\diff{x'}F_{\sigma}(x)F_{\sigma}(x')
        \hat{\pi}(x,t)\hat{\pi}(x',t')\proj{0}{0} +\text{H.c.}\right).
    \end{align}
Using the smeared Wightman function $W_{\sigma}[t,t']$, we can rewrite the above:
    \begin{align} \label{eqn:nosubrho}
        \nonumber
        \hat\rho_{\text{q},\text{out}}^{(2)} &= - \lambda^2 \int_{-\infty}^{\infty}\diff{t} \int_{-\infty}^{\infty}\diff{t'} \chi(t)\chi(t') \hat{\mu}(t)\hat\rho_{\text{q},0}\hat{\mu}(t')W_{\sigma}[t',t] \\
        -\lambda^2 &\int_{-\infty}^{\infty}\diff{t}\int_{-\infty}^t\diff{t'}
        \chi(t)\chi(t')\hat{\mu}(t)\hat{\mu}(t')\hat\rho_{\text{q},0}W_{\sigma}[t,t'] +\text{H.c.},
    \end{align}
where
    \begin{align}
        W_{\sigma}[t,t'] =\! \int\diff{x}\!\!\!\!\int\diff{x'}\!\!F_{\sigma}(x)F_{\sigma}(x') \trpi \left[ \hat{\pi}(x,t)\ket{0}\bra{0}\hat{\pi}(x',t') \right].
    \end{align}
Since the state of the field is initially pure, we can rewrite the term $\trpi \left[ \hat{\pi}(x,t)\ket{0}\bra{0}\hat{\pi}(x',t') \right]$ as the expectation value of $\hat\pi(x,t)\hat\pi(x',t')$ on the state $\ket{0}$:
    \begin{align}
        W_{\sigma}[t,&t'] = 
        \frac{1}{4\pi}
        \int\diff{x}
        \int\diff{x'}
        F_{\sigma}(x)  F_{\sigma}(x')					\\ \nonumber
        &	\int\diff{k}\!\! \int\diff{k'}
        C_{\varepsilon}(k) C_{\varepsilon}(k')	\sqrt{|k|} \sqrt{|k'|} \\
        & \bra{0}(\hat{a}_k^{\dagger} 	\ee^{\ii(|k|t-kx)} 		- \text{H.c.})
        (\hat{a}_{k'}^{\dagger} \ee^{\ii(|k'|t'-k'x')} 	- \text{H.c.})
        \ket{0}.
        \nonumber
    \end{align}
Using that \mbox{$\bra{0}\hat{a}_k\hat{a}_{k'}^{\dagger}\ket{0} = \delta(k-k')$}, we arrive at:
    \begin{align}
        W_{\sigma}[t,t'] &= 
        \frac{1}{4\pi}
        \int\diff{k}
        \int\diff{x} 	F_{\sigma}(x)e^{\ii kx}\\ \nonumber
        &\int\diff{x'} 	F_{\sigma}(x')e^{-\ii kx'}
        C_{\varepsilon}^2(k)	|k| 	\ee^{-\ii|k|(t-t')}.
    \end{align}
We can write this more concisely as
    \begin{align}\label{eqn:wightman}
        W_{\sigma}[t,t'] &= \frac{1}{4\pi} \int\diff{k}
        \tilde{F}_{\sigma}^2(k) C_{\varepsilon}^2(k) |k| \ee^{-\ii |k|(t-t')},
    \end{align}
where $\tilde F_{\sigma} (k)$ is the Fourier transform of the spatial distribution, i.e., 
    \begin{equation}
        \tilde{F}_{\sigma}(k) = \int\diff{x}F_{\sigma}(x)\ee^{\ii k x}.
    \end{equation} 
Note also that our choice of real symmetric smearing means that $\tilde{F}_{\sigma}(k)=\tilde{F}_{\sigma}(-k)$.

We can now write~\eqref{eqn:nosubrho} in a more compact fashion:
    \begin{align}\label{rhodt2a}
        \nonumber
        \hat\rho_{\text{q},\text{out}}^{(2)} &= 
        \int_{-\infty}^{\infty}\!\!\!\diff{t} 
        \int_{-\infty}^{\infty}\!\!\!\diff{t'}
        \chi(t) \chi(t') W_{\sigma}[t',t]   \ee^{ \ii\Omega(t-t')}\ket{e}\bra{e} \\ 
        -& 2 
        \int_{-\infty}^{\infty}\!\!\!\!	\diff{t} \!\!\!
        \int_{-\infty}^{t}\!\!\!\!		\diff{t'}
        \chi(t) \chi(t') Re[W_{\sigma}[t,t']]\cos(\Omega(t-t'))\ket{g}\bra{g}.
    \end{align}    
Substituting~\eqref{eqn:wightman} in, we finally obtain an expression for the final state of the qubit that is general in choice of shape, cutoff, and switching functions, as in equations~\eqref{eqn:rhoout} and~\eqref{eqn:pe}:
    \begin{align}\label{rhodt2g}
        \nonumber
        \hat\rho_{\text{q},\text{out}}^{(2)} &= \frac{1}{4\pi}
        \int\diff{k}  \tilde{F}_{\sigma}^2(k) C_{\varepsilon}^2(k)	|k|	
        \left(\:
        \int_{-\infty}^{\infty}\!\!\! \diff{t}\!\!\!
        \int_{-\infty}^{\infty}\!\!\! \diff{t'} 
        \chi(t) \chi(t')\right. \\ \nonumber &\ee^{\ii(|k|+\Omega)(t-t')}\ket{e}\bra{e}
        - 2
        \int_{-\infty}^{\infty}\!\!\! \diff{t} \!\!\!
        \int_{-\infty}^{t}\!\!\!		\diff{t'}
        \chi(t) \chi(t') \\
        & \cos(-|k|(t-t')) 
        \cos(\Omega(t-t'))\ket{g}\bra{g}
        \big).
    \end{align}
    
\bibliography{bibliography}
\end{document}